\documentclass[twocolumn,showpacs,preprintnumbers,amsmath,amssymb,prl]{revtex4}
\usepackage{graphicx}
\usepackage{amsmath}
\usepackage{dcolumn}% Align table columns on decimal point
\usepackage{bm}% bold math

\newcommand{\Hamil}{{\cal H}}
\newcommand{\kk}{{\mathbf k}}
\newcommand{\pp}{{\mathbf p}}
\newcommand{\RR}{{\mathbf R}}

\begin{document}

\title{Indications of coherence-incoherence crossover in layered transport}

\author{Urban Lundin}
 \email{lundin@physics.uq.edu.au}
\author{Ross H.\ McKenzie}
\affiliation{Department of Physics, University of Queensland,
             Brisbane Qld 4072, Australia}

\date{\today}

\begin{abstract}
For many layered metals the temperature
dependence of the interlayer resistance has a different
behavior than the intralayer resistance.
In order to better understand interlayer transport
we consider a concrete model which exhibits this behavior.
A small polaron model is used to illustrate how the
interlayer transport is related to the coherence
of quasi-particles within the layers.
Explicit results are given for
the electron spectral function, interlayer optical conductivity and the 
interlayer magnetoresistance. All these quantities have two contributions: 
one coherent (dominant at low temperatures) and one 
incoherent (dominant at high temperatures). 

\end{abstract}
\maketitle

Many of the most interesting strongly correlated
electron materials have a layered crystal structure
and highly anisotropic electronic properties.
Examples include the cuprates~\cite{ando96,lavrov98,hussey},
colossal magneto-resistance 
materials~\cite{kimura96,takenaka99}, organic molecular
crystals~\cite{mihaly00,buravov}, 
and strontium ruthenate~\cite{hussey98}.
One particularly poorly understood property
of these materials is that often the resistivity
perpendicular to the layers has quite a distinct
temperature dependence to that parallel to the layers~\cite{millis02}.
This is in contrast to what is expected for
an anisotropic Fermi liquid: the 
parallel and perpendicular resistivity then have the same
temperature dependence, being determined by
the intralayer scattering rate, $\Gamma(T)$.
(This result holds even when $\Gamma(T)$          
is larger than the interlayer bandwidth~\cite{moses}).
In many of these materials the interlayer
resistivity is a non-monotonic function of temperature
with a maximum at some temperature 
$ T_{\perp}^{\mathrm{max}} $.
In some of the materials the intralayer resistivity
also has a maximum as a function of temperature,
but at a higher temperature $ T_{\parallel}^{\mathrm{max}} > 
 T_{\perp}^{\mathrm{max}} $.
Previously 
%{\bf INSERT REF}
 it has been suggested that the
maximum in the temperature dependence of the
interlayer resistivity is associated with a crossover
from coherent interlayer transport at low temperatures
to incoherent transport at high temperatures.
There is no consensus as to what actually determines
$T_{\perp}^{\mathrm{max}}$.
Recent angle resolved photoemission spectroscopy (ARPES)
experiments on two different layered cobalt oxide compounds~\cite{valla02}
found that one only observed peaks in the
electronic spectral function
(corresponding to coherent quasi-particle excitations
within the layers) below a temperature $T^{\mathrm{coh}}$
that was comparable to $T_{\perp}^{\mathrm{max}}$.

Although many theoretical papers have considered
the problem of incoherent interlayer transport
(see for example Ref. \onlinecite{biermann}
and the references given in Ref.~\onlinecite{moses}),
 we are unaware on any theory
which starts with a many-body Hamiltonian
and produces the three temperature scales,
$T^{\mathrm{coh}}$,  $T_{\perp}^{\mathrm{max}}$,
and $T_{\parallel}^{\mathrm{max}}$. 
In this paper we consider a simple microscopic
model which exhibits
$T^{\mathrm{coh}}  \sim T_{\perp}^{\mathrm{max}}
< T_{\parallel}^{\mathrm{max}}$. 
The model is a layered version of Holstein's 
molecular crystal model where the electrons
strongly couple to bosonic excitations
to produce small polarons.
It should be stressed that we are not claiming that the
charge transport involves small polarons in all of the above materials .
Rather, we suggest that this model can 
provide insight into the relevant physics
associated with these temperature scales.

%We consider the simplest possible model which will clearly show a explicit 
%crossover, the application to materials mentioned here 
%In this paper we will demonstrate that a model for polaronic transport have 
%these features of a coherent and an incoherent contribution and at different 
%temperatures the properties of the system is dominated by one of the two 
%contributions. However, application to the materials mentioned above
%might not be straightforward. Layered transport has been 
%investigated theoretically before~\cite{johansson94}, but we will focus 
%on the temperature dependence and the crossover. 
%Yaun and Thalmeier~\cite{yuan99} considered the charge ordering in the layered 
%compound LaSr$_2$Mn$_2$O$_7$ as caused by polaron formation. 
%We extend the idea presented by Alexandrov and Bratkovsky~\cite{alexandrov99}, 
%and apply it to layered transport. 

\begin{figure}[hb]
\includegraphics*[scale=1.0]{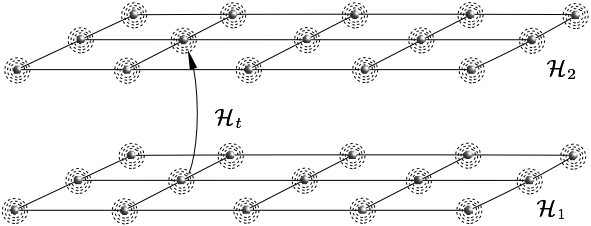}
\caption{The two layers with polarons are coupled by a hopping Hamiltonian.}
\label{layers:fig}
\end{figure}

We start with a Holstein model~\cite{holstein59} 
 for an infinite layered system
where the electrons interact with dispersionless bosons.
The Hamiltonian is
\begin{eqnarray}
{\cal H}&=&\sum_i\epsilon^0c^{\dag}_ic_i+
           \hbar\omega_0\sum_i a^{\dag}_ia_i+
           \sum_{<i\eta>}t_{i\eta}c^{\dag}_{\eta}c_i \nonumber \\
         &&+M\sum_{i}c^{\dag}_ic_i (a_i+a^{\dag}_i), \nonumber
%        &&+ \sum_{i,\qq}M_{\qq}c^{\dag}_ic_ie^{i\qq\cdot\RR_i}
%               (a_{\qq}+a^{\dag}_{-\qq}), \nonumber
\end{eqnarray}
where $\epsilon^0$ is the on-site energy, $\omega_0$ is the 
characteristic frequency of the bosons, $t_{i\eta}$ is the 
hopping integral  
between nearest-neighbor
 sites $i$ and $\eta$, $M$ is the coupling between the bosons and 
the electrons. We introduce a dimensionless coupling 
$g=\left(\frac{M}{\hbar\omega_0}\right)^2$. We have that $g\agt 1$ 
in order for small polaronic effects to be important. 
Since we want to study layered systems we split the hopping
into parallel and perpendicular to the layers,
 $t_{\parallel}$ and $t_{\perp}$ respectively  
where $t_{\parallel}\gg t_{\perp}$. This enables us to write the 
Hamiltonian in a way more adapted for the layered case, shown 
in Fig.~\ref{layers:fig}. 
The nature of the transport depends on how $t_{\parallel}$ and $t_{\perp}$ 
compares with $\Gamma$, the scattering rate due to the bosons. 
We assume that that $\Gamma > t_{\perp}$, so that the inter-layer transport 
can be described by considering two decoupled layers.  
The Hamiltonian can be specified for this system. 
Two layers are coupled with a hopping Hamiltonian. Within each layer 
the electrons can hop but there is a coupling to a bosonic degree of freedom
in each layer with characteristic frequency $\omega_0$.
The bosons can be phonons, magnons, plasmons, or particle-hole excitations.   
We only consider a single frequency $\omega_0$ for reasons of simplicity;
it allows us to express some of our results in a analytical form.

We then use the Hamiltonian: 
\begin{equation}
{\cal H}={\cal H}_1+{\cal H}_2+{\cal H}_t \nonumber
\end{equation}
where  
\begin{eqnarray}
{\cal H}_1&=&\sum_i\epsilon^0c^{\dag}_ic_i+
           \hbar\omega_0\sum_ia^{\dag}_ia_i+
           t_{\parallel}\sum_{<i\eta>}c^{\dag}_{\eta}c_i \nonumber \\
         &&+M\sum_ic^{\dag}_ic_i(a_i+a^{\dag}_i), \nonumber \\
%        &&+ \sum_{i,\qq}M_{\qq}c^{\dag}_ic_ie^{i\qq\cdot\RR_i}
%               (a_{\qq}+a^{\dag}_{-\qq}), \nonumber \\
{\cal H}_2&=&\sum_j\epsilon^0d^{\dag}_jd_j+
           \hbar\omega_0\sum_jb^{\dag}_jb_j+
           t_{\parallel}\sum_{<j\delta>}d^{\dag}_{\delta}d_j \nonumber \\
          &&+M\sum_jd^{\dag}_jd_j(b_j+b^{\dag}_j), \nonumber \\
%         &&+ \sum_{j,\pp}M_{\pp}d^{\dag}_jd_je^{i\qq\cdot\RR_j}
%               (b_{\pp}+b^{\dag}_{-\pp}), \nonumber \\
{\cal H}_t&=&t_{\perp}\sum_{i}(c^{\dag}_id_i+{\rm h.c.}). \nonumber
\end{eqnarray}
Here, and below, $c,i,a,1$ refers to one layer, and 
$d,j,b,2$ to the other one.
First we focus on the properties of the two separate layers.            
We perform a Lang-Firsov transformation~\cite{lang63} 
to remove the coupling of the electrons to the bosons.
Then $c_i \to \tilde{c}_i = c_i X_i$ 
where 
\begin{equation}
X_i=\exp\left[\frac{M}{\hbar\omega_0}(a_i-a_i^{\dag})\right],
\label{X-op:eq}
\end{equation}
and
$a_i \to a_i - {M \over \omega_0} c_i^\dagger c_i$.
The Hamiltonian is transformed to 
$\bar{{\cal H}}=e^S{\cal H}e^{-S}$
where $S=\frac{M}{\hbar\omega_0}\sum_i c_i^\dagger c_i(a_i^\dagger -a_i)$.
 This diagonalizes the electron-boson part of the Hamiltonian, 
but introduces extra $X$-operators~\cite{mahan} 
in the hopping parts of the Hamiltonian. 
For example: 
\begin{equation}
\bar{\Hamil}_t=t_{\perp}\sum_{<ij>}
(c^{\dag}_iX^{\dag}_id_jY_j+{\rm h.c.}),
\label{barHt:eq}
\end{equation}
The intralayer hopping terms can be treated by adding and subtracting
to the Hamiltonian  a term
\begin{equation}
t_\parallel \sum_{<ij>}
 \langle X_i X^{\dag}_j\rangle 
c^{\dag}_ic_j 
= \sum_{\kk} \epsilon_{\kk} 
c^{\dag}_{\kk}c_{\kk} 
\end{equation}
%where ${\sigma}_{ij}= \langle T_{\tau} X_i(0)X^{\dag}_j(0)\rangle,$ 
where $\langle  ..\rangle$ denotes a thermal average over boson states
and this term describes a tight-binding band of small polarons
within each layer~\cite{lang63,alexandrov}
\begin{equation}
\epsilon_{\kk} = \epsilon^0
- e^{-g(1+2n_B)}
t_\parallel [ \cos(k_x a  ) + \cos(k_y a ) ], 
\label{tight:eq}
\end{equation}
where $a$ is the lattice constant within the layers and
$n_B(T) = (\exp( \hbar \omega_0/ k_B T) -1 )^{-1}$ is the Bose function. 

There is then a residual interaction~\cite{alexandrov}
between the polarons and the bosons which is described by        
\begin{equation}
\bar{\Hamil}_{p-b}=
t_\parallel \sum_{<ij>}
%[\hat{\sigma}_{ij}-\sigma_{ij}]
 [X_i X^{\dag}_j - \langle X_i X^{\dag}_j\rangle ] 
                       c^{\dag}_ic_j, 
\label{interaction:eq}
\end{equation}
and leads to scattering of the small polarons.
The first non-zero contribution to the imaginary part of the 
polaron self-energy, 
$\Sigma$, comes when the polaron emits one boson and absorbs one boson. 
If the energy dependence of the
the  density of states (DOS) is neglected one finds that~\cite{alexandrov}.
\begin{equation}
\Gamma(T)=Wg^2n_B(T)(1+n_B(T)), 
\label{decay:eq}
\end{equation} 
where $W$ is the renormalized bandwidth,
$W=4 t_\parallel e^{-g(1+2n_B)}$~\cite{lang63,alexandrov,mahan}. 

Note that the small polarons are composite particles
(quasi-particles). They consist of an electron bound
to a ``cloud'' of bosons. This coherent quantum state can
move freely within the layers producing coherent charge transport.
In contrast, 
ARPES involves ejection of electrons rather than polarons from
the crystal. Similarly, the interlayer charge transport
involves the tunneling of electrons between layers.
In order for this to occur the bosons bound to the electron
in the polaron must be removed, the electron tunnels, and
a new set of bosons are bound to the electron.

{\it Greens function within a single layer.}
Let us start by calculating the Matsubara Green function (GF) within one 
layer, ignoring the coupling between the layers. 
After the Lang-Firsov transformation we calculate the small polaron GF
\begin{eqnarray}
G^0(\kk,\tau)&=&-i\Theta(\tau)\frac{1}{N}\sum_{i,i'}e^{i\kk\cdot(\RR_{i'}-\RR_i)}
                \langle T_{\tau}
              \tilde{c}_{i'}(\tau)\tilde{c}^{\dag}_i(0)\rangle 
 \nonumber \\
&=&-i\Theta(\tau) e^{(\epsilon_{\kk}- i \Gamma) i\tau/h}, \nonumber
\end{eqnarray}
where $\Theta(\tau)$ is the step function,
$T_{\tau}$ is the time ordering operator,
$\RR_i$ is the position of lattice site $i$.
%$\epsilon_{\kk}$ is the dispersion of the polarons that 
%is subject to narrowing~\cite{lang63,alexandrov,mahan},
$\Gamma$ is the inverse lifetime associated 
with the interaction, Eq.(\ref{interaction:eq}),
and there are $N$ lattice sites in each layer. 

The {\em electron} GF, $G(\kk,i\omega_n)$ involves a convolution
of the polaron GF with the Fourier transformed (denoted ${\cal F}$) 
$X$-operators~\cite{alexandrov2}
% with the $\hat{\sigma}_{ij}$-term 
\begin{equation}
G(\kk,i\omega_n)=
\frac{1}{N}\sum_{\omega_{n'},i,i',\kk'}
{\cal F}\{ \langle X_{i'}(\tau) X^{\dag}_i(0)\rangle \}
%\sigma(\RR_m,\omega_{n'}-\omega_n)
           G^0(\kk',\omega_{n'})e^{i(\kk-\kk')\cdot (\RR_{i'}-\RR_i)}. \nonumber
\end{equation}
%Making some simplifications we come to the following expression 
%\begin{eqnarray}
%&&
%G(\kk,i\omega_n)=
%e^{-N^{-1}\sum_{\qq}|\gamma_{\qq}|^2(1+2n_B)}\frac{1}{N}
%\sum_{\RR_M,\kk}e^{i(\kk-\kk')\cdot\RR_m} \nonumber \\
%&&
%\sum_{l=-\infty}^{\infty} 
%\frac{I_l[N^{-1}\sum_{\qq}|\gamma_{\qq}|^2\cos(\qq\cdot\RR_m)
%           \sqrt{n_B(1+n_B)}]e^{-l\hbar\omega_0\beta/2}}
%     {i\omega_n-\epsilon_{\kk'}+l\hbar\omega_0+i\Gamma}.
%\end{eqnarray}
%We assume a square lattice and that $\epsilon_{\kk}$ is given 
%by a tight binding expression,
%where  $\RR_m=\RR_i-\RR_j$ and 
%\begin{equation}
%\sigma( \RR_i-\RR_j,\tau) = 
% <X_i(\tau) X^{\dag}_j(0)> 
%\end{equation}
Performing the summation over $\RR_{i'}-\RR_i$ we obtain                
\begin{eqnarray}
&&G(\kk,i\omega_n)=e^{-g(1+2n_B)}
\left\{
       \frac{1}{i\omega_n-\epsilon_{\kk}+i\Gamma} \right. \nonumber \\
&&+
       \sum_{\kk'}\frac{I_0\left[2g\sqrt{n_B(1+n_B)}\right]-1}
                       {i\omega_n-\epsilon_{\kk'}+i\Gamma} \nonumber \\
&&\left.+\sum_{\kk',l\neq 0}
            \frac{I_{l}\left[2g\sqrt{n_B(1+n_B)}\right]
                  e^{-l\hbar\omega_0\beta/2}}
                 {i\omega_n-\epsilon_{\kk'}+l\hbar\omega_0+i\Gamma}
\right\}. 
\label{GF:eq}
\end{eqnarray}
% We introduced a decay, $\Gamma$, 
%which comes from the polaron-boson interaction. 

Note that the GF is a sum of a coherent and a incoherent part, 
i.e., the terms on the second and third lines are {\em independent} of $\kk$. 
Eq.(\ref{GF:eq}) is a generalization to non-zero temperatures 
of the GF found by Alexandrov
and Ranninger~\cite{alexandrov2}. 

\begin{figure}
\includegraphics{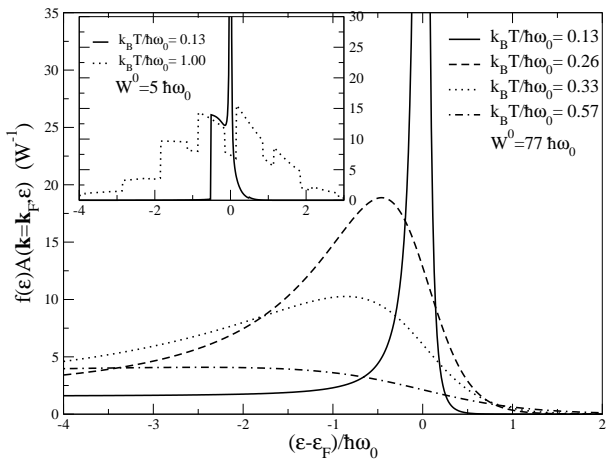}
\caption{Energy dependence of the
electron spectral function at a wave-vector on the Fermi surface, using 
a constant DOS.
The product of the spectral function $A(\kk_F,\epsilon)$
with the Fermi-Dirac distribution function $f(\epsilon)$
is shown because this can be compared with ARPES spectra.
Note that the well-defined quasi-particle peak which occurs for
$k_BT\ll \hbar\omega_0$ disappears at higher temperature.
The results are shown for an electron-boson coupling of
$g=1$ . The inset shows the same thing for a smaller bandwidth.}
\label{spectral:fig}
\end{figure}

In Fig.~\ref{spectral:fig} we plot the electron spectral function, 
$A(\kk,\omega)=\mathrm{Im}[G(\kk,\omega)]$,  
resulting from the GF, Eq.(\ref{GF:eq}), for different temperatures.
With increasing temperature the boson modes become populated,
$n_B(T)$ increases and the spectral weight shifts from
the coherent part of the spectral function
to the incoherent part.

Qualitatively similar behavior was seen in recent 
ARPES~\cite{valla02} measurements. 
This behavior does not change much qualitatively when $g$ is changed. 
From plots we estimated that the crossover takes place at 
%From Eq.(\ref{GF:eq}) we can estimate when the crossover in the spectral 
%function takes place. By ignoring the third line we would get that 
\begin{equation}
k_BT^{\mathrm{coh}}\sim\frac{\hbar\omega_0}{2g}. 
\end{equation}
This can also be justified using Eq.(\ref{GF:eq}) when $W<\hbar\omega_0$. 
%{\bf THIS MIGHT BE CHANGED} 

{\it Interlayer conductivity.}                  
Let us derive an expression for the current perpendicular to the planes. 
%The current operator for a field in the perpendicular direction 
%(from L to R) is 
%\begin{equation}
%\hat{j}_{LR}=\frac{ie}{\hbar}\sum_{i,j}t_{\perp}({\bf R}_i-{\bf R}_j)
%d^{\dag}_jc_iY^{\dag}_{j}X_i. 
%\end{equation}
We assume that the hopping between the layers only takes place between nearest 
neighbors (see Fig.~\ref{layers:fig}). 
At applied field $V$ the current is given by 
\begin{equation}
I_{\mu\nu}(eV)=\frac{2e}{h}\mathrm{Im}\left[\int_0^{\beta} d\tau^{ieV\tau}
\langle T_\tau \hat{j}_{\mu}(\tau)\hat{j}^{\dag}_{\nu}(0)\rangle\right],
\label{currcurr:eq}
\end{equation}
where $\hat{j}$ is the current operator. $\mu$ and $\nu$ are 
directions in the crystal. 

%We consider one site at a time and can therefore ignore the sum over $i$ in 
%the current operator. 
%The total conductivity will be proportional to the 

Using the Hamiltonian above and Eq.(\ref{currcurr:eq}) we have,  
to second order in $t_{\perp}$,
\begin{eqnarray}
&&\hspace*{-5mm}
I_{\perp}(eV)= \frac{2e}{h}t_{\perp}^2d^2\sum_{j,j_1} 
               \int_0^{\beta} d\tau e^{ieV\tau} 
\nonumber \\
&&\times
\left\langle 
T_\tau c^{\dag}_j(\tau)d_{j_1}(\tau)
 d^{\dag}_j(0)c_{j_1}(0)\right\rangle
\left\langle T_\tau Y^{\dag}_{j_1}(\tau)Y_j(\tau)
 X^{\dag}_j(0)X_{j_1}(0)\right\rangle, \nonumber
\label{2decop:eq}
\end{eqnarray}
where $d$ is the distance between the 2 layers. 

We assume that the bosons 
in separate layers are independent of one another.
Hence, in Eq.(\ref{2decop:eq}), we can 
decouple the $X$ and $Y$ polaron operators corresponding
to the first and second layers. 
This means that the Fourier transformed averages of the electron operators 
gives rise to two GFs. These GFs describe polarons bands within each layer. 

%\begin{eqnarray}
%&&\left\langle T c^{\dag}_j(t)c_{j_1}(0)\right\rangle \rightarrow 
%               G_L(p,ip_n), \nonumber \\
%&&\left\langle T d_{j_1}(t)d^{\dag}_j(0)\right\rangle \rightarrow 
%               G_R(k,ip_n-i\omega). \nonumber
%\end{eqnarray}
%The average of the polaron-operators can be written as an exponential 
%$\langle T X^{\dag}_j(\tau)X_{j_1}(0)\rangle=e^{\Phi_{jj_1}(\tau)}$~\cite{lang63,mahan}. 
%\begin{eqnarray}
%&&\hspace*{-8mm}\langle T X^{\dag}_j(\tau)X_{j_1}(0)\rangle
%                \langle T Y^{\dag}_{j_1}(\tau)Y_j(0)\rangle=
%   e^{2\Phi_{jj_1}(\tau)}=
%%e^{-N^{-1}\sum_{\qq}|\gamma_{\qq}|^2(1/2+n_B)} \nonumber \\
%%&&\times\exp\left\{(2N)^{-1}\sum_{\qq}|\gamma_{\qq}|^2
%%                   \cos[\qq\cdot(\mathbf{R}_j-\mathbf{R}_{j_1})]
%%                   [(1+n_B)e^{\omega \tau} +n_Be^{-\omega \tau}]\right\} \nonumber\\
%e^{-2g(1+2n_B)} \nonumber \\
%&&\times\prod_{\qq}\sum_{l=-\infty}^{\infty} I_l\left[4g
%    \cos[\qq\cdot(\mathbf{R}_j-\mathbf{R}_{j_1})]
%    \sqrt{n_B(1+n_B)}\right]e^{il\omega_0(\tau+i\beta/2)}. \nonumber 
%\label{expo:eq}
%\end{eqnarray}
%$I_l$ is a modified Bessel function of order $l$. 
%$|\gamma_{\qq}|^2=1/2\left(\frac{M_{\qq}}{\hbar\omega_{\qq}}\right)^2$. 
%We can perform the Fourier transform in $\omega$, and if we assume 
%that the electron GFs has a imaginary part, $\Gamma$, which is taken from 
%Eq.(\ref{decay:eq}), we get the following for the current 
%Performing a Fourier transform 
After some algebra we can write the current as
\begin{widetext}
\begin{eqnarray}
&&I_{\perp}(eV)=\frac{2e}{h}t_{\perp}^2d^2e^{-2g(1+2n_B)}%\nonumber \\
  \left\{\int_{-\infty}^{\infty}\frac{d\epsilon}{2\pi} 
         \sum_{\kk}A^0_1(\kk,\epsilon)A^0_2(\kk,\epsilon+eV)
        \left[f(\epsilon)-f(\epsilon+eV)\right]\right. \nonumber \\
&&\left. +\left(I_0\left[4g\sqrt{n_B(1+n_B)}\right]-1\right)
          \int_{-\infty}^{\infty}\frac{d\epsilon}{2\pi} 
          \sum_{\kk}A^0_1(\kk,\epsilon)\sum_{\pp}A^0_2(\pp,\epsilon+eV)
          \left[f(\epsilon)-f(\epsilon+eV)\right]\right. \nonumber \\
&&\left. +\sum_{l\neq 0 \atop l=-\infty}^{\infty}
          I_{l}\left[4g\sqrt{n_B(1+n_B)}\right]
          e^{-l\hbar\omega_0\beta/2}
          \int_{-\infty}^{\infty}\frac{d\epsilon}{2\pi} 
          \sum_{\kk}A^0_1(\kk,\epsilon)\sum_{\pp}
                        A^0_2(\pp,\epsilon+eV+l\hbar\omega_0)
 \left[f(\epsilon)-f(\epsilon+eV+l\hbar\omega_0)\right]\right\}
\nonumber \\ 
\label{current:eq}
\end{eqnarray}
\end{widetext}
$\pp$ belongs to the first (1) layer and $\kk$ to the second (2) layer. 
$A^0_1$ and $A^0_2$ are the spectral functions for the polaron 
GFs in each layer respectively, and
$I_l$ is a modified Bessel function of order $l$. 

Note the similarity in structure between Eq.(\ref{GF:eq}) and 
Eq.(\ref{current:eq}).
The expression for the current, Eq.(\ref{current:eq}), has a contribution 
from coherent and one from incoherent transport. 
The first line corresponds to tunneling where the momentum of the polaron 
parallel to the layers is conserved. In the second and third
terms the  intralayer 
momentum is not conserved. The third line has a
difference energy of $l\hbar\omega_0$ between the
polarons in the two layers because there is
a non-zero difference  
between    the  net number of bosons that are
 absorbed and emitted in the tunneling event. 
At low temperature the coherent part dominates but at high temperature 
($k_BT>\hbar\omega_0$) the incoherent mechanism 
of transport will dominate. Thus, there is a {\em crossover} from coherent to 
incoherent transport. 

Then, the inter-layer conductivity is easily obtained 
$\sigma_{\perp}=e\left.\frac{dI_{\perp}}{d(eV)}\right|_{eV=0}$. 
This should be multiplied with the number of sites in one layer. 
The interlayer resistivity, measured in experiments, 
is simply $\sigma_{\perp}^{-1}$. 

By using the fact that when $\Gamma\ll W$, 
$\sum_{\kk}A(\kk,\epsilon)=D(\epsilon)$, where $D(\epsilon)$ is the DOS, 
and assuming that terms containing the derivative of the 
spectral function is small $dA(\epsilon+eV)/dV\ll A(\epsilon+eV)D(\epsilon)$
We can write the conductivity as:
\begin{widetext}
\begin{eqnarray}
&&\sigma_{\perp}=\frac{2e^2}{h}t_{\perp}^2d^2e^{-2g(1+2n_B)}
  \left\{\int_{-W/2}^{W/2}\frac{d\epsilon}{2\pi} 
         D(\epsilon)\left[\frac{1}{2\Gamma}+
         D(\epsilon)\left(I_0\left[4g\sqrt{n_B(1+n_B)}\right]-1\right)\right]
         \frac{-df(\epsilon)}{d\epsilon}\right. \nonumber \\
&&\left. +\sum_{l\neq 0 \atop l=-\infty}^{\infty}
          I_{l}\left[4g\sqrt{n_B(1+n_B)}\right]
          e^{-l\hbar\omega_0\beta/2}
          \int_{-\infty}^{\infty}\frac{d\epsilon}{2\pi} 
          D(\epsilon)D(\epsilon+l\hbar\omega_0)
          \frac{-df(\epsilon+l\hbar\omega_0)}{d\epsilon}\right\}
\nonumber \\ 
\label{conductivity:eq}
\end{eqnarray}
\end{widetext}

%Equivalently, we can work directly with
%electron operators and we find
In general,  the interlayer conductivity 
for decoupled layers is~\cite{moses},
\begin{equation}
\sigma_{\perp}=\frac{2e^2}{h}t_{\perp}^2\int d\epsilon 
               \sum_{\kk}A^2(\kk,\epsilon)\left[-\frac{df}{d\epsilon}\right],
\label{currcurr2:eq}
\end{equation}
where $A(\kk,\epsilon)$ is the {\em electron} spectral
function for a single layer.
Directly substituting Eq.(\ref{GF:eq}) in this we can
obtain Eq.(\ref{conductivity:eq}).

\begin{figure}
\includegraphics{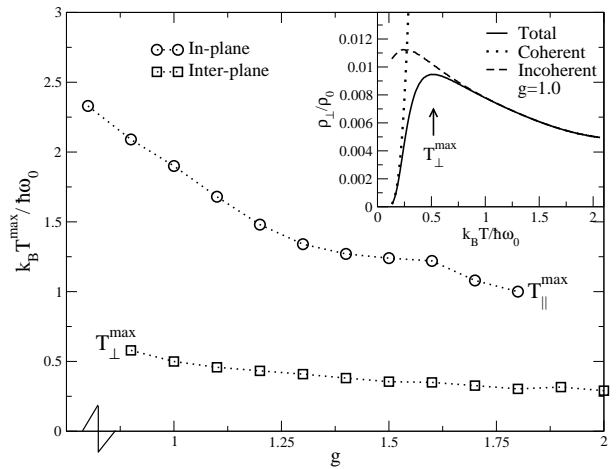}
\caption{Peak temperature in the resistivity as a function of 
coupling. Results obtained 
using a tight binding DOS (no major change was seen when using a constant 
DOS). The two sets of data-points are for the intralayer and the interlayer 
crossover temperatures respectively. The intralayer crossover occurs at much 
higher temperatures than the interlayer. 
The inset shows the interlayer 
conductivity as a function of temperature when $g=1$ and $W^0=77\hbar\omega_0$. 
$\rho_0=h/\left[2e^2d^2\left(\frac{t_{\perp}}{t_{\parallel}}\right)^2\right]$.}
\label{crossover:fig}
\end{figure}

We can estimate the temperature of the crossover in the conductivity by 
having equal 
contribution from the coherent and the incoherent parts. If we look at 
the conductivity there is a minima (maxima in the resistivity), 
corresponding to the crossover 
(see inset in Fig.~\ref{crossover:fig}). Ignoring the 
contribution from the $l\neq 0$ terms in Eq.(\ref{current:eq}) 
we can get an approximate expression 
for the crossover temperature for the interlayer resistivity:
\begin{equation}
k_BT^{\mathrm{max}}_{\perp}
\sim\frac{\hbar\omega_0}{\sqrt[4]{2^3}g}
\sim 0.59 \frac{\hbar\omega_0}{g}.
\label{Tmax:eq}
\end{equation}
This expression compares quite well to the crossover temperature extracted 
from a numerical plot of the resistivity versus temperature 
in Fig.~\ref{crossover:fig}. 
Hence, we see that $T^{\mathrm{coh}}$ and $T^{\mathrm{max}}_{\perp}$
are comparable. At these temperatures, $\Gamma \sim 0.3 W$
A more detailed analysis shows that the $l\neq 0$ terms in 
Eq.(\ref{current:eq}) contributes linearly in $W$. So the estimate 
for the crossover, Eq.(\ref{Tmax:eq}), should only hold when 
$W\alt \hbar\omega_0$, but extensive numerical work showed that it has a wide 
range of validity. 

It is interesting to note that 
$\sigma^{\mathrm{coh}}\sim \sigma^{\mathrm{incoh}}$ actually 
occurs at a temperature {\em lower} than $T^{\mathrm{max}}_{\perp}$, 
as can be seen in the inset in Fig.~\ref{crossover:fig}. 
This is because an almost constant contribution from the $l\neq 0$ terms. 

{\it Intralayer conductivity.}
When we calculated the current between the layers we assumed that 
the small polaron band was well-defined.
However, it is known that even within the layer, 
as the temperature increases, there is a crossover
from coherent to incoherent transport, occurring at 
$T^{\mathrm{max}}_{\parallel}$, 
and there is a maximum in the resistivity
associated with this crossover~\cite{holstein59}.
An important question is the size
of $T^{\mathrm{max}}_{\parallel}$ relative to
$T^{\mathrm{max}}_{\perp}$.

We are unaware of a systematic method of calculating
the full temperature dependence of the intralayer
conductivity. At low temperatures it can
be calculated by considering the band transport
of local polarons. At high temperatures
one can consider the diffusion (or hopping)
of localized polarons. Extrapolating the results
to intermediate temperatures provides a means
to estimate the crossover temperature and the
regions of validity of the different expressions.

% If the conductivity within the 
%layers is coherent, we have well defined quasi-particles and the current 
%between the layers is as above.
% Otherwise we have to start with localized polarons.
% Let us have a look at low and high temperatures~\cite{holstein59}. 

At low temperatures 
($\Gamma \ll W$)
 the transport in the layers 
are coherent. We assume that 
we have well developed quasi-particles in the layers and the bosons acts as 
a small perturbation. The conductivity is given by: 
\begin{equation}
\sigma_{\parallel}=\frac{e^2}{2\pi^2} \int d \epsilon \int d^2k {v(\kk)^2     
\over \Gamma}
\delta(\epsilon - \epsilon_\kk)
                   \left(-\frac{df}{d\epsilon}\right). \nonumber
\end{equation}
We use a tight binding model for the electrons to calculate the velocities, 
and the decay $\Gamma$ is taken from Eq. (\ref{decay:eq}) above. 
%{\bf ROSS CONNECTION TO HOLSTEIN IS NOT STRAIGHT FORWRD, HE CALCULATES 
% DIFFUSIVITY, WE DO IT IN ANOTHER WAY, BETTER TO DISCUSS PHYSICS, NOT 
% RELATION TO HOLSTEINS CALCULATION} 
%Using the tight binding approximation for the dispersion we get:
%\begin{equation}
%\sigma_{\parallel}=\frac{2e^2}{h}
%\frac{2\beta \tilde{t_{\parallel}}}{g^2n_B(1+n_B)}
%\int_{-2\pi}^{2\pi}dxdy
%\frac{\sin^2(x)}
%     {1+\cosh[\beta(\epsilon_0+\tilde{t}\cos(x)+\tilde{t}\cos(y)-\mu)]},
%\end{equation}
%where $\tilde{t_{\parallel}}$ is the renormalized hopping parameter 
%$t_{\parallel}e^{-g(1+2n_B)}$.

At high temperatures 
($\Gamma \gg W$)
 the electron is trapped by the formation of the polaron. 
The electrons are localized at the lattice sites and
the concept of a wave-vector for the small polaron is meaningless. 
The intralayer hopping term in the Hamiltonian should then 
be treated as the perturbation. 
%The electron GF is $G=(\omega-\Delta+\Sigma+i\Gamma)^{-1}$. 
%$\Delta=\epsilon_0-g\hbar\omega_0<0$. 
The intralayer conductivity at high temperatures 
from Eq.(\ref{currcurr:eq}) depends on the polaron GF, solved when the 
hopping term is the perturbation.
However, due to the lack of dispersion of the 
bosons the time integral in the conductivity
diverges~\cite{mahan}. We overcome this problem
by including a finite self energy in the electron GF. 
To evaluate this we sum a series of the lowest order diagrams. 
We have to take special care about the number of paths on the 2D lattice. 
If $G(\omega)$ is the local electron GF it has a self energy that satisfies
\begin{equation}
\Sigma(\omega)
%=4\sum_{n=1}^{\infty}[1+2n(1-n)]t_\parallel^{2n}G^{2n-1}
= 4t_\parallel^2G\frac{1+t_\parallel^2G^2+t_\parallel^4G^4}
{(1-t_\parallel^2G^2)^3}. 
\end{equation}
%The 4 $X$-operators are decoupled~\cite{mahan}. The result is Fourier 
%transformed and we have the current. 
%\begin{equation}
%I_{\parallel}=\frac{2e}{h}t_{\parallel}^2e^{-2g(1+2n_B)}
%\mathrm{Im}\left\{
%\int_{-\infty}^{\infty} \mathrm{d}\omega G(\omega)\left[
%\sum_{l=-\infty}^{\infty}I_l\left[4g\sqrt{n_B(1+n_B)}\right]
%e^{-l\hbar\omega_0\beta/2} G(eV+\omega+l\hbar\omega_0)-G(eV+\omega)
%\right]\right\}. 
%\end{equation}
This is a self-consistent equation for the real and imaginary part of 
the self energy that have to be solved simultaneously to self consistency. 
This was done on a square lattice where the polaron can hop to four different 
sites each time. The conductivity for the low and high temperature regions 
were plotted and the maxima in $\rho_{\parallel}$ was extracted. 

There is a crossover also for the intralayer resistivity. 
From Fig.~\ref{crossover:fig}, the crossover occurs roughly when 
\begin{equation}
k_BT^{\mathrm{max}}_{\parallel}\sim 2\frac{\hbar\omega_0}{g},
\end{equation}
using this in Eq.(\ref{decay:eq}), means that the crossover occurs when
$\Gamma\sim 8W$. 
The intralayer crossover occurs at higher temperatures than the interlayer 
crossover, therefore the assumption made above that for the interlayer 
calculation we have well developed quasi-particles within each
layer is justified. 

{\it Optical conductivity.}
The frequency dependence of the interlayer optical conductivity, 
$\sigma_\perp(\omega)$, 
has been suggested to be a probe of interlayer coherence
in the metallic state~\cite{katsufuji}.
The optical conductivity can be found from
Eq.(\ref{currcurr:eq}), by letting $eV\rightarrow \omega$, and calculating 
the derivative.

\begin{figure}
\includegraphics{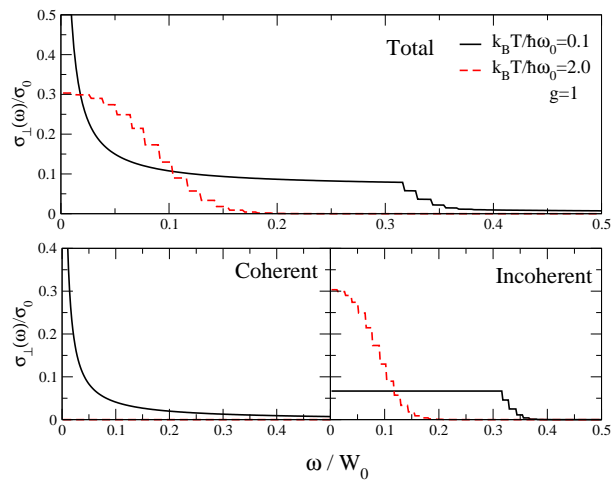}
\caption{Optical conductivity divided into the two contributions, coherent and 
incoherent, plotted for two different temperatures. In the lower left 
panel we plot the coherent part, and in the right the incoherent part.
$\sigma_0=2e^2d^2\left(\frac{t_{\perp}}{t_{\parallel}}\right)^2/h$.}
\label{opt_cond:fig}
\end{figure}

Fig.~\ref{opt_cond:fig} shows how at low temperatures there is a well-defined
Drude peak at zero frequency due to coherent interlayer transport of 
small polarons. The width of this feature is approximately $\Gamma$.
Note that this feature occurs even though $\Gamma > t_{\perp}$,
as has been pointed out previously~\cite{moses2}.
As the temperature increases the spectral weight of this
feature decreases and is replaced
with a broader feature associated with incoherent
interlayer transport and with a width that is determined
by the small polaron bandwidth {\it within} the layers.
The incoherent part becomes narrower with increasing 
temperature because of the polaron narrowing of the bands. 
Changing $g$ and $t_\parallel$ does not qualitatively
change this temperature-dependent crossover.

{\it Magnetoresistance.}
If we apply a magnetic field, $B$, parallel to the layers 
(the x-y plane) we have a 
orbital effect on the 
paths of the electrons. This can be described by a shift in
the Bloch wave vector,
$\kk\rightarrow\kk-\frac{e}{\hbar}\mathbf{A}$, 
where $\mathbf{A}$ is the vector potential for the magnetic field. 
For a magnetic field in the $x$ direction, when an electron
tunnels between adjacent layers it undergoes a shift in
the y-component of its wave vector by $-d B$~\cite{moses}.
In the general expression Eq.(\ref{currcurr2:eq}) $A^2(\kk,\epsilon)$ 
is replaced
with $A(\kk,\epsilon)A(\kk + \frac{e}{\hbar} d B \vec{y})$.
However, since the incoherent part of the conductivity contains a summation 
over $\kk$-space 
and is {\em independent} of $\kk$, this will be unaffected by the 
magnetic field. 
Thus, we will have two contributions to the interlayer
conductivity and one is $B$-independent: 
\begin{equation}
\sigma_{\perp}(B)=\sigma_{\perp}^{\mathrm{coh}}(B)+
                  \sigma_{\perp}^{\mathrm{incoh}}(B=0).
\label{twopart}
\end{equation}
$\sigma^{\mathrm{coh}}(B)$ decreases 
with increasing magnetic field~\cite{schofield,moses}
\begin{equation}
\sigma_{\perp}^{\mathrm{coh}}(B)=\frac{\sigma_{\perp}^{\mathrm{coh}}(B=0)}
             {\sqrt{1+(e v_F c B \Gamma)^2}}.
\end{equation}
where $v_F$ is the Fermi velocity.
If we increase $B$, the coherent part decreases, and, therefore, 
$T_{\perp}^{\mathrm{max}}$ would shift to lower values. 
A separation of the conductivity in two parts,
as in Eq.(\ref{twopart}), has been proposed previously
on a phenomenological basis, in order to describe
the magnetoresistance of Sr$_2$RuO$_4$~\cite{hussey98}.
(Except there a weak field dependence is associated
with the incoherent contribution due to 
Zeeman splitting).

We have shown that a small polaron model for transport
 in layered systems shows a 
crossover from coherent to incoherent transport 
at different temperatures for intralayer and interlayer 
transport. It can be observed in ARPES, as well as in measurements of 
magnetoresistance and optical conductivity. 
It is sometimes suggested (or assumed) 
that the maximum in the interlayer resistivity as
a function of temperature occurs at a temperature
$ T_{\perp}^{\mathrm{max}} $ 
determined by the strength of the interlayer
hopping $t_\perp$, either by 
$k_B T_{\perp}^{\mathrm{max}} \sim t_\perp$ or 
$ \Gamma(T_{\perp}^{\mathrm{max}}) \sim t_\perp$
where $ \Gamma(T)$ is the temperature dependent
scattering rate within the layers.
However, we find that $T_{\perp}^{\mathrm{max}}$ can occur
at a higher temperature, which is actually
independent of $t_\perp$, and instead closely
related to $T^{\mathrm{coh}}$.

\begin{acknowledgments}
U.\ Lundin acknowledges the support from the Swedish foundation for
international cooperation in research and higher education (STINT). 
This work was also supported by the Australian Research Council (ARC).
After completion of this work, we became aware 
of some related work by Ho and Schofield concerning a small 
polaron model for interlayer transport. 
We thank them for sending us a copy of their preprint prior to submission.
\end{acknowledgments}

\bibliography{paper}

\end{document}